\documentclass[aps,pra,showpacs,superscriptaddress,preprint]{revtex4}
\usepackage{amssymb}
\usepackage{amsmath}
\usepackage{graphicx}

\catcode`ð=\active
 \defð{\u{g}}
 \catcode`Ð=\active
 \defÐ{\u{G}}
 \catcode`Ý=\active
\defÝ{\. I}
 \catcode`ö=\active
\defö{\"{o}}
 \catcode`Ö=\active
 \defÖ{\"O}
 \catcode`ü=\active
 \defü{\"{u}}
 \catcode`Ü=\active
 \defÜ{\"{U}}
 \catcode`Þ=\active
\defÞ{\c{S}}
 \catcode`þ=\active
 \defþ{\c{s}}
 \catcode`ý=\active
 \defý{{\i}}
 \catcode`ç=\active
\defç{\d{c}}
 \catcode`Ç=\active
\defÇ{\d{C}}

\input{tcilatex}

\begin{document}

\title{Effects of external fields on two-dimensional Klein-Gordon particle
under pseudoharmonic oscillator interaction}
\author{Sameer M. Ikhdair}
\email[E-mail: ]{sikhdair@neu.edu.tr; sikhdair@gmail.com}
\affiliation{Physics Department, Near East University, 922022 Nicosia, North Cyprus,
Turkey}
\author{Majid Hamzavi}
\email[E-mail: ]{ majid.hamzavi@gmail.com (Corresponding author)}
\affiliation{Department of Basic Sciences, Shahrood Branch, Islamic Azad University,
Shahrood, Iran}
\date{%
\today%
}

\begin{abstract}
We study the effects of the perpendicular magnetic and Aharonov-Bohm (AB)
flux fields on the energy levels of a two-dimensional (2D) Klein-Gordon (KG)
particle subjects to equal scalar and vector pseudo-harmonic oscillator
(PHO). We calculate the exact energy eigenvalues and normalized wave
functions in terms of chemical potential parameter, magnetic field strength,
AB flux field and magnetic quantum number by means of the Nikiforov-Uvarov
(NU) method. The non-relativistic limit, PHO and harmonic oscillator
solutions in the existence and absence of external fields are also obtained.

\end{abstract}

\maketitle

\newpage

\section{Introduction}

The pseudo-harmonic oscillator (PHO) potential is extensively used to
describe the bound state of the interaction systems and has been applied for
both classical and modern physics. It plays a basic role in chemical and
molecular physics since it can be used to calculate the molecular
vibration-rotation energy spectrum of linear and non-linear systems [1-3].
This potential is considered as an intermediate between harmonic oscillator
(HO) and Morse-type potentials which are more realistic anharmonic
potentials. In the non-relativistic quantum mechanics, the PHO is one of the
exactly solvable potentials in the frame of the Schr\"{o}dinger equation and
has also been studied in one-dimensional (1D), two-dimensional (2D),
three-dimensional (3D) and even in D-dimensional space.

Therefore, this problem has attracted a great deal of interests in solving
the Schr\"{o}dinger and Klein-Gordon (KG) equations with the PHO
interaction. The discrete (bound) and continuous (scattering) energy spectra
of the PHO have been investigated by the SU(1,1) spectrum generating algebra
[4]. The PHO potential has been obtained depending on the dimension and the
angular momentum and the zero-point energy in two-dimension was found to be
minimum. The exact polynomial solution of the Schr\"{o}dinger equation for
PHO has been obtained in 3D space [5]. Recently, solutions of the Schr\"{o}%
dinger and KG equations with PHO potential have been investigated [6-10]. A
realization of the creation and annihilation (ladder) operators for the
solution to the Schr\"{o}dinger equation with a PHO in 2D was studied in
Refs. [11,12]. The operators satisfy the commutation relations of an SU(1,1)
group. The exact solution of the Schr\"{o}dinger equation with a PHO in an
arbitrary dimension D was presented in Ref. [13]. The exact analytical
solutions of the Schr\"{o}dinger equation of the D-dimansional space for the
PHO potential have been presented by means of the ansatz method and the
energy eigenvalues were calculated from the eigenfunction ansatz [14]. The
exact bound-state solutions of the KG and the Dirac equations with equal
scalar and vector PHO potential have been obtained using the supersymmetric
quantum mechanics, shape invariance and other alternative methods [15]. The
bound-state solutions of the Dirac equation with PHO have been obtained in
the presence of spin and pseudospin symmetries [16]. In relativistic quantum
mechanics, the treatment of this problem is relatively inadequate.

Recently, several works have been achieved in this direction for which the
solutions of non-relativistic and relativistic equations with different
potentials for their bound and continuum states were investigated. The
approximate analytical solutions of the scattering states were derived for $%
D $-dimensional Schr\"{o}dinger equation with modified P\"{o}schl-Teller,
Eckart, modified Morse and generalized Hulth\'{e}n potentials were found for
any angular momentum states using new approximation scheme to centrifugal
term [17-20]. It was shown that the energy levels of the continuum states of
modified Morse potential reduce to those of bound states at the poles of the
scattering amplitude [19]. Using the supersymmetry and shape-invariance,
exact bound state solutions of Schr\"{o}dinger equation with P\"{o}%
schl-Teller double-ring-shaped Coulomb potential were presented [21]. Exact
solution of the one-dimensional KG equation with mixed scalar and vector
linear potentials were studied in the context of deformed quantum mechanics
characterized by a finite minimal uncertainty in position using the momentum
space representation [22]. New ring-shaped harmonic oscillator for spin-$1/2$
particles was studied and corresponding eigenfunctions and eigenvalues were
obtained by using the Dirac equation with equal mixture of scalar and vector
potentials [23]. Further, the approximate analytical solution of the Dirac
equation with the P\"{o}schl-Teller potential has been presented for
arbitrary spin-orbit quantum number in view of spin-symmetry [24].

Recently, the spectral properties in a 2D charged particle (electron or
hole) confined by a PHO potential in the presence of external strong uniform
magnetic field $\overrightarrow{B}$ along the $z$ direction and
Aharonov-Bohm (AB) flux field created by a solenoid have been studied. The
Schr\"{o}dinger equation is solved exactly for its bound states (energy
spectrum and wave functions) [25,26]. So, it is natural that the
relativistic effects for a charged particle under the action of this
potential could become important, especially for a strong coupling.

The aim of the present work is to study the exact analytical bound state
energy eigenvalues and normalized wave functions of the spinless
relativistic equation with equal scalar and vector PHO interaction under the
effect of external uniform magnetic field and AB flux field in the framework
of the NU method [27,28]. The non-relativistic limit of our solution is
obtained by making an appropriate mapping of parameters. Further, the KG-PHO
and KG-HO special cases are also treated.

The structure of this paper is as follows. We study the effect of external
uniform magnetic and AB flux fields on a relativistic spinless particle
(anti-particle) under equal scalar and vector PHO interaction in Section 2.
We discuss some special cases in Section 3. Finally, we give our concluding
remarks in Section 4.

\section{KG Equation with PHO Interaction Under External Fields}

The KG equation is a wave equation mostly used in describing particle
dynamics in relativistic quantum mechanics. Nonetheless, physically this
equation describes a scalar particle (spin $0$). Moreover, this wave
equation, for free particles, is constructed using two objects: the
four-vector linear momentum operator $P_{\mu }=i\hbar \partial _{\mu }$ and
the scalar rest mass $M,$ allows one to introduce naturally two types of
potential couplings. One is the gauge-invariant coupling to the four-vector
potential $\left\{ A_{\mu }\left( \overrightarrow{r}\right) \right\} _{\mu
=0}^{3}$ which is introduced via the minimal substitution $P_{\mu
}\rightarrow P_{\mu }-gA_{\mu },$ where $g$ is a real coupling parameter.
The other, is an additional coupling to the space-time scalar potential $S_{%
\text{conf}}(\overrightarrow{r})$ which is introduced by the substitution $%
M\rightarrow M+S_{\text{conf}}(\overrightarrow{r}).$ The term
\textquotedblleft four-vector\textquotedblright\ and \textquotedblleft
scalar\textquotedblright\ refers to the corresponding unitary irreducible
representation of the Poincar$\mathbf{{\acute{e}}}$ space-time symmetry
group (the group of rotations and translations in ($3+1$)-dimensional
Minkowski space-time). Gauge invariance of the vector coupling allows for
the freedom to fix the gauge (eliminating the non physical gauge modes)
without altering the physical content of the problem. Many choose to
simplify the solution of the problem by taking the space component of the
vector potential to vanish (i.e., $\overrightarrow{A}$). One may write the
time-component of the four-vector potential as $gA_{0}=V_{\text{conf}}(\vec{r%
}),$ then it ends up with two independent potential functions in the KG
equation. These are the \textquotedblleft vector\textquotedblright\ $V_{%
\text{conf}}(\overrightarrow{r})$ and the \textquotedblleft
scalar\textquotedblright\ $S_{\text{conf}}(\overrightarrow{r})$ potentials
[29,30]$.$

The free KG equation is written as 
\begin{equation}
(\partial ^{\mu }\partial _{\mu }+M^{2}c^{4})\psi _{KG}(t,\overrightarrow{r}%
)=0.
\end{equation}%
Moreover, the vector and scalar couplings mentioned above introduce
potential interactions by mapping the free KG equation for a 2D single
charged electron as 
\begin{equation}
\left[ c^{2}\left( \overrightarrow{p}+\frac{e}{c}\overrightarrow{A}\right)
^{2}-\left( E-V_{\text{conf}}(\vec{r})\right) ^{2}+\left( Mc^{2}+S_{\text{%
conf}}(\vec{r})\right) ^{2}\right] \psi (\overrightarrow{r},\phi )=0,
\end{equation}%
where we used the transformation $\overrightarrow{p}\rightarrow 
\overrightarrow{p}+\frac{e}{c}\overrightarrow{A}.$ Further, the $2D$
cylindrical wave function $\psi (\overrightarrow{r},\phi )$ is defined as%
\begin{equation}
\psi (\vec{r},\phi )=\frac{1}{\sqrt{2\pi }}e^{im\phi }g(r),\text{ }m=0,\pm
1,\pm 2,\ldots ,
\end{equation}%
where $m$ is the magnetic quantum number. This type of coupling attracted a
lot of attention in the literature due to the resulting simplification in
the solution of the relativistic problem. The scalar-like potential coupling
is added to the scalar mass so that in case when $S_{\text{conf}}(\vec{r}%
)=\pm V_{\text{conf}}(\overrightarrow{r}),$ the KG equation could always be
reduced to a Schr\"{o}dinger-type second order differential equation as
follows%
\begin{equation}
\left[ c^{2}\left( \overrightarrow{p}+\frac{e}{c}\overrightarrow{A}\right)
^{2}+2\left( E\pm Mc^{2}\right) V_{\text{conf}}(\overrightarrow{r}%
)+M^{2}c^{4}-E^{2}\right] \psi (\overrightarrow{r},\phi )=0.
\end{equation}%
The potential $V_{\text{conf}}(\overrightarrow{r})$ is taken as the
repulsive PHO potential [1-3,11]: 
\begin{equation}
V_{\text{conf}}(\vec{r})=V_{0}\left( \frac{r}{r_{0}}-\frac{r_{0}}{r}\right)
^{2},
\end{equation}%
where $r_{0}$ and $V_{0}$ are the zero point (effective radius) and the
chemical potential, respectively. Further, the vector potential $%
\overrightarrow{A}$ may be represented as a sum of two terms, $%
\overrightarrow{A}=\overrightarrow{A}_{1}+\overrightarrow{A}_{2}$ such that $%
\overrightarrow{\nabla }\times \overrightarrow{A}_{1}=\overrightarrow{B}$
and $\overrightarrow{\nabla }\times \overrightarrow{A}_{2}=0,$ where $%
\overrightarrow{B}$ $=B\widehat{z}$ is the applied magnetic field, and $%
\overrightarrow{A}_{2}$ describes the additional Aharonov-Bohm (AB) flux
field $\Phi _{AB}$ created by a solenoid in cylindrical coordinates [31].
Hence, the vector potentials have the following azimuthal components [32] 
\begin{equation}
\overrightarrow{A}_{1}=\frac{Br}{2}\widehat{\phi },\text{ }\overrightarrow{A}%
_{2}=\frac{\Phi _{AB}}{2\pi r}\widehat{\phi },\text{ }\overrightarrow{A}%
=\left( \frac{Br}{2}+\frac{\Phi _{AB}}{2\pi r}\right) \widehat{\phi }.
\end{equation}%
Hence, the bound-state solutions of the two cases in Eq. (4) are to be
treated separately as follows.

\subsection{The bound states for positive energy}

The positive energy states (corresponding to $S_{\text{conf}}(%
\overrightarrow{r})=+V_{\text{conf}}(\overrightarrow{r})$ in the
non-relativistic limit (taking $E-Mc^{2}\rightarrow E$ and $\left(
E+Mc^{2}\right) \rightarrow 2\mu c^{2},$ where $M$ is an effective mass and $%
\left\vert E\right\vert \ll Mc^{2}$) are solutions of%
\begin{equation}
\left[ \frac{1}{2\mu }\left( \overrightarrow{p}+\frac{e}{c}\overrightarrow{A}%
\right) ^{2}+2V_{\text{conf}}(\overrightarrow{r})-E\right] \psi (%
\overrightarrow{r},\phi )=0,
\end{equation}%
where $\psi (\overrightarrow{r},\phi )$ stands for either $\psi ^{(+)}(%
\overrightarrow{r},\phi )$ or $\psi ^{(\text{KG})}(\overrightarrow{r},\phi
). $ This is the Schr\"{o}dinger equation for the potential $2V_{\text{conf}%
}(\overrightarrow{r}).$ Thus, the choice $S_{\text{conf}}(\overrightarrow{r}%
)=+V_{\text{conf}}(\overrightarrow{r})$ produces a nontrivial
non-relativistic limit with a potential function $2V_{\text{conf}}(%
\overrightarrow{r}),$ and not $V_{\text{conf}}(\overrightarrow{r}).$
Accordingly, it would be natural to scale the potential term in Eq. (4) and
Eq. (7) so that in the non-relativistic limit the interaction potential
becomes $V_{\text{conf}},$ not $2V_{\text{conf}}.$ Thus, we need to recast
Eq. (4) for the $S(\vec{r})=V(\vec{r})$ as [29,33] 
\begin{equation}
\left[ c^{2}\left( -i\hbar \overrightarrow{\nabla }+\frac{e}{c}%
\overrightarrow{A}\right) ^{2}+2\left( E+Mc^{2}\right) V_{\text{conf}}(\vec{r%
})\right] \psi (\overrightarrow{r},\phi )=\left( E^{2}-M^{2}c^{4}\right)
\psi (\overrightarrow{r},\phi ),
\end{equation}%
and in order to simplify Eq. (8) we introduce new parameters $\lambda
_{1}=E+Mc^{2}$ and $\lambda _{2}=E-Mc^{2}$ so that it can be reduced to the
form%
\begin{equation}
\left[ c^{2}\left( -i\hbar \overrightarrow{\nabla }+\frac{e}{c}%
\overrightarrow{A}\right) ^{2}-\lambda _{1}\left( \lambda _{2}-V_{\text{conf}%
}(\overrightarrow{r})\right) \right] \psi (\overrightarrow{r},\phi )=0.
\end{equation}%
Now, inserting Eqs. (3), (5) and (6) into the KG equation (9) and further
introducing a change of variable $s=r^{2},$ that maps $r\in (0,\infty )$ to s%
$\in (0,\infty ),$ we obtain second-order differential equation satisfying
the radial wave function $g(s),$ 
\begin{equation}
g^{\prime \prime }(s)+\frac{1}{s}g^{\prime }(s)+\frac{1}{s^{2}}\left( -\frac{%
\gamma ^{2}}{4}s^{2}+\frac{\nu ^{2}}{4}s-\frac{\beta ^{2}}{4}\right) g(s)=0,%
\text{ }g(0)=0\text{ and }g(\infty )=0,
\end{equation}%
where we have employed the following abbreviation symbols 
\begin{subequations}
\begin{equation}
\nu ^{2}=\frac{1}{\hbar ^{2}c^{2}}\left[ \lambda _{1}\left( \lambda
_{2}+2V_{0}\right) -Mc^{2}\omega _{c}\hbar m^{\prime }\right] ,
\end{equation}%
\begin{equation}
\beta ^{2}=m^{\prime 2}+\frac{1}{\hbar ^{2}c^{2}}r_{0}^{2}V_{0}\lambda _{1},%
\text{ }m^{\prime }=m+\xi ,\text{ }m^{\prime }=1,2,\ldots ,
\end{equation}%
\begin{equation}
\gamma ^{2}=\left( \frac{M\omega _{c}}{2\hbar }\right) ^{2}+\frac{1}{\hbar
^{2}c^{2}}\frac{V_{0}\lambda _{1}}{r_{0}^{2}},
\end{equation}%
with $m^{\prime }$ is a new quantum number. Here $\xi =\Phi _{AB}/\Phi _{0}$
is taken as integer with the flux quantum $\Phi _{0}=hc/e,$ $\omega
_{c}=eB/Mc$ is the cyclotron frequency$.$ Now, we use the basic ideas of the
NU method [27] and the parametric NU derived in Ref. [28] to obtain energy
spectrum equation 
\end{subequations}
\begin{equation}
\nu ^{2}=2\left( 2n+1+\beta \right) \gamma ,\text{ }n=0,1,2,\ldots .
\end{equation}%
where the constant parameters used in our calculations are displayed in
Table 1. Inserting Eqs. (11a)-(11c), we finally arrive at the following
transcendental energy formula,%
\begin{equation*}
2\left( 1+2n+\sqrt{m^{\prime }{}^{2}+\frac{V_{0}r_{0}^{2}\lambda _{1}}{\hbar
^{2}c^{2}}}\right) \sqrt{\left( \frac{M\omega _{c}}{2\hbar }\right) ^{2}+%
\frac{V_{0}\lambda _{1}}{\hbar ^{2}c^{2}r_{0}^{2}}\text{ }}
\end{equation*}%
\begin{equation}
=\frac{1}{\hbar ^{2}c^{2}}\left[ \lambda _{1}\left( \lambda
_{2}+2V_{0}\right) -Mc^{2}\hbar \omega _{c}m^{\prime }\right] ,\text{ }%
m^{\prime }=1,2,\ldots ,
\end{equation}%
We may find solution to the above transcendental equation as $%
E=E_{KG}^{(+)}. $ In the non-relativistic $\lim $it when $\lambda
_{1}\rightarrow 2Mc^{2}$ and $\lambda _{2}\rightarrow E,$ the above energy
equation has a solution given by Eq. (39) of Ref. [25].

Using Eq. (38) of Ref. [28] and Table 1, we find the corresponding radial
wave function $g(r)$ as%
\begin{equation}
g(r)=C_{n,m}r^{\left\vert \beta \right\vert }e^{-\gamma r^{2}/2}F\left(
-n,\left\vert \beta \right\vert +1;\gamma r^{2}\right) ,
\end{equation}%
and hence the total $2D$ KG wave function (3) takes the explicit form%
\begin{equation}
\psi _{n,m}^{(+)}(\vec{r},\phi )=\frac{1}{\sqrt{2\pi }}e^{im\phi }\sqrt{%
\frac{\gamma ^{\left\vert \beta \right\vert +1}n!}{\pi \left( n+\left\vert
\beta \right\vert \right) !}}r^{\left\vert \beta \right\vert }e^{-\gamma
r^{2}/2}L_{n}^{(\beta )}(\gamma r^{2}),
\end{equation}%
where $L_{a}^{\left( b\right) }\left( x\right) =\frac{\left( a+b\right) !}{%
a!b!}F\left( -a,b+1;x\right) $ is the associated Laguerre polynomial and $%
F(-a,b;x)$ is the confluent hypergeometric function. Note that the wave
function (15) is finite and satisfying the standard asymptotic analysis for $%
r=0$ and $r\rightarrow \infty .$

\subsection{The bound states for negative energy}

When $S_{\text{conf}}(\overrightarrow{r})=-V_{\text{conf}}(\overrightarrow{r}%
)$, we need to follow same procedure of solution in subsection A and
consider the solution given by (12) with the changes 
\begin{subequations}
\begin{equation}
\nu ^{2}\rightarrow \widetilde{\nu }^{2}=\frac{1}{\hbar ^{2}c^{2}}\left[
\lambda _{2}\left( \lambda _{1}+2V_{0}\right) -Mc^{2}\omega _{c}\hbar
m^{\prime }\right] ,
\end{equation}%
\begin{equation}
\beta ^{2}\rightarrow \widetilde{\beta }^{2}=m^{\prime }{}^{2}+\frac{1}{%
\hbar ^{2}c^{2}}r_{0}^{2}V_{0}\lambda _{2},
\end{equation}%
\begin{equation}
\gamma ^{2}\rightarrow \widetilde{\gamma }^{2}=\left( \frac{M\omega _{c}}{%
2\hbar }\right) ^{2}+\frac{1}{\hbar ^{2}c^{2}}\frac{V_{0}\lambda _{2}}{%
r_{0}^{2}}.
\end{equation}%
Hence, the negative energy solution for antiparticle can be readily found as 
\end{subequations}
\begin{equation*}
2\left( 2n+1+\sqrt{m^{\prime }{}^{2}+\frac{1}{\hbar ^{2}c^{2}}%
r_{0}^{2}V_{0}\lambda _{2}}\right) \sqrt{\left( \frac{M\omega _{c}}{2\hbar }%
\right) ^{2}+\frac{1}{\hbar ^{2}c^{2}}\frac{V_{0}\lambda _{2}}{r_{0}^{2}}}
\end{equation*}%
\begin{equation}
=\frac{1}{\hbar ^{2}c^{2}}\left[ \lambda _{2}\left( \lambda
_{1}+2V_{0}\right) -Mc^{2}\hbar \omega _{c}m^{\prime }\right] ,\text{ }%
m^{\prime }=1,2,\ldots ,
\end{equation}%
and the $2D$ wave function is%
\begin{equation}
\psi _{n,m}^{(-)}(\vec{r},\phi )=\sqrt{\frac{\widetilde{\gamma }^{\left\vert 
\widetilde{\beta }\right\vert +1}n!}{\pi \left( n+\left\vert \widetilde{%
\beta }\right\vert \right) !}}r^{\left\vert \widetilde{\beta }\right\vert
}e^{-\widetilde{\gamma }r^{2}/2}L_{n}^{(\widetilde{\beta })}(\widetilde{%
\gamma }r^{2})\frac{1}{\sqrt{2\pi }}e^{im\phi }.
\end{equation}%
It should be noted that the negative energy states are free fields since
under these conditions Eq. (4) can be rewritten as%
\begin{equation}
\left[ -\frac{1}{2\mu }\left( \overrightarrow{p}+\frac{e}{c}\overrightarrow{A%
}\right) ^{2}+E\right] \psi _{n,m}(\overrightarrow{r},\phi )=0,
\end{equation}%
which is a simple free-interaction mode. Further, the set of parameters
given in Eqs. (16a)-(16c) becomes%
\begin{equation}
\widetilde{\nu }=\sqrt{\frac{2ME}{\hbar ^{2}}-\frac{M\omega _{c}}{\hbar }%
m^{\prime }},\text{ }\widetilde{\beta }=m^{\prime },\text{ }\widetilde{%
\gamma }=\frac{M\omega _{c}}{2\hbar }.
\end{equation}%
Thus, Eq. (17) gives the following energy formula%
\begin{equation}
E_{nm^{\prime }}^{(-)}=\left( n+m^{\prime }+\frac{1}{2}\right) \hbar \omega
_{c},
\end{equation}%
and hence the wave function reads%
\begin{equation}
\psi _{nm^{\prime }}^{(-)}(\vec{r},\phi )=\sqrt{\frac{\left( \frac{M\omega
_{c}}{2\hbar }\right) ^{m^{\prime }+1}n!}{\pi \left( n+m^{\prime }\right) !}}%
r^{m^{\prime }}e^{-\frac{M\omega _{c}}{4\hbar c}r^{2}}L_{n}^{(m^{\prime
})}\left( \frac{M\omega _{c}}{2\hbar }r^{2}\right) e^{im\phi }.
\end{equation}

\section{Discussions}

In this section we briefly study some special cases and relationship between
our results and some other authors':

\subsection{Schr\"{o}dinger-PHO problem under the effect of magnetic and AB
flux fields}

In the non-relativistic limit ($\lambda _{1}\rightarrow 2M$ and $\lambda
_{2}\rightarrow E$), Eq. (13) can be easily reduced into the form: 
\begin{subequations}
\begin{equation}
E_{nm}(\xi ,\beta )=\hbar \Omega \left( n+\frac{\left\vert \widetilde{m}%
\right\vert +1}{2}\right) +\frac{1}{2}\hbar \omega _{c}m^{\prime }-2V_{0},
\end{equation}%
\begin{equation}
\text{ }\Omega =\sqrt{\omega _{c}^{2}+4\omega _{D}^{2}},\text{ }\omega _{D}=%
\sqrt{2V_{0}/Mr_{0}^{2}},\text{ }\left\vert \widetilde{m}\right\vert =\sqrt{%
m^{\prime 2}+a^{2}},
\end{equation}%
where $a=k_{F}r_{0}$ with $k_{F}=\sqrt{2MV_{0}/\hbar ^{2}}$ is the fermi
wave vector of the electron [25]. Further, the wave function becomes

\end{subequations}
\begin{equation}
\psi _{n,m}^{(+)}(\vec{r},\phi )=\frac{1}{\sqrt{2\pi }}e^{im\phi }\sqrt{%
\frac{c^{b+1}n!}{\pi \left( n+b\right) !}}%
r^{b}e^{-cr^{2}/2}L_{n}^{(b)}(cr^{2}),
\end{equation}%
with the given parameters%
\begin{equation*}
b=\sqrt{m^{\prime 2}+\frac{2MV_{0}r_{0}^{2}}{\hbar ^{2}}}\text{ and \ }c=%
\sqrt{\left( \frac{M\omega _{c}}{2\hbar }\right) ^{2}+\frac{2MV_{0}}{\hbar
^{2}r_{0}^{2}}}.
\end{equation*}

\subsection{KG-PHO problem}

The energy spectrum of relativistic spinless particle in the absence of
magnetic and AB flux fields has the form%
\begin{equation}
\frac{2\hbar }{r_{0}}\sqrt{\text{ }V_{0}}\left( 1+2n+\sqrt{m{}^{2}+\frac{%
V_{0}r_{0}^{2}}{\hbar ^{2}}\lambda _{1}}\right) =\left( \lambda
_{2}+2V_{0}\right) \sqrt{\lambda _{1}},
\end{equation}%
and is reduced to its non-relativistic limit:

\begin{equation}
E_{nm}=-2V_{0}+\left( 1+2n+\sqrt{m{}^{2}+\frac{2MV_{0}r_{0}^{2}}{\hbar ^{2}}}%
\right) \sqrt{\text{ }\frac{2V_{0}\hbar ^{2}}{Mr_{0}^{2}}},
\end{equation}%
which is completely identical to Eq. (27) of Ref. [7]. The wave function can
be expressed as

\begin{equation}
\psi _{n,m}^{(+)}(\vec{r},\phi )=\frac{1}{\sqrt{2\pi }}e^{im\phi }\sqrt{%
\frac{C^{B+1}n!}{\pi \left( n+B\right) !}}%
r^{B}e^{-Cr^{2}/2}L_{n}^{(B)}(Cr^{2}),
\end{equation}%
with 
\begin{equation*}
B=\sqrt{m^{2}+\frac{r_{0}^{2}V_{0}}{\hbar ^{2}}\lambda _{1}}\text{ \ and \ }%
C=\frac{1}{\hbar r_{0}}\sqrt{V_{0}\lambda _{1}}.
\end{equation*}

\subsection{KG-HO problem}

The energy equation of the relativistic spinless particle subject to the HO
field is%
\begin{equation}
n^{\prime }\hbar c\sqrt{\text{ 2k}}-\sqrt{\lambda _{1}}\lambda _{2}=0,\text{ 
}n^{\prime }=1,2,\ldots ,
\end{equation}%
where $n^{\prime }=1+\left\vert m\right\vert +2n,$ $n=0,1,2,\ldots ,$ which
is completely identical to Eq. (11) and Eq. (26) of Ref. [34] when one uses
the notation $k=2V_{0}/r_{0}^{2}=M\omega _{D}^{\prime 2}.$ Following Ref.
[35], the Eq. (28) has three solutions, the only real solution giving energy
is%
\begin{equation}
E_{nm}=\frac{1}{3}\left( Mc^{2}+M^{2}c^{4}T^{-1/3}+T^{1/3}\right) ,
\end{equation}%
with%
\begin{equation}
T=27kn^{\prime 2}\hbar ^{2}c^{2}-8M^{3}c^{6}+3n^{\prime }\hbar c\sqrt{%
3k\left( 27kn^{\prime 2}\hbar ^{2}c^{2}-16M^{3}c^{6}\right) }.
\end{equation}%
The wave function takes the form

\begin{equation}
\psi _{n,m}^{(+)}(\vec{r},\phi )=\sqrt{\frac{D^{\left\vert m\right\vert +1}n!%
}{\pi \left( n+\left\vert m\right\vert \right) !}}r^{\left\vert m\right\vert
}e^{-Dr^{2}/2}L_{n}^{(\left\vert m\right\vert )}(Dr^{2})e^{im\phi },\text{ }%
D=\frac{1}{\hbar r_{0}}\sqrt{V_{0}\lambda _{1}}.
\end{equation}%
If one expands Eq. (28) as a series of $\lambda _{2},$ it becomes 
\begin{equation}
n^{\prime }\hbar =\sqrt{\frac{M}{k}}\left[ \lambda _{2}+\frac{1}{4Mc^{2}}%
\lambda _{2}^{2}-\frac{1}{32M^{2}c^{4}}\lambda _{2}^{3}+O(\lambda _{2}^{4})%
\right] ,
\end{equation}%
and taking the first order of $\lambda _{2}$ by neglecting the higher order
relativistic corrections, we finally arrive at the non-relativistic
solution: 
\begin{equation}
E_{nm}^{\prime }=E_{nm}-Mc^{2}=\hbar \sqrt{\text{ }\frac{k}{M}}\left(
1+2n+\left\vert m\right\vert \right) =\left( 1+\left\vert m\right\vert
+2n\right) \hbar \omega _{D}^{\prime },\text{ }n=0,1,2,\ldots ,
\end{equation}%
and wave function resembles the one given in (31) with $D=\frac{1}{\hbar
r_{0}}\sqrt{2MV_{0}}.$

\section{Concluding Remarks}

In this work, we have obtained bound state energies and wave functions of
the relativistic spinless particle subject to a PHO interaction and
expressed in terms of external uniform magnetic field and AB flux field. We
explored the solution of both positive (particle) and negative
(anti-particle) KG energy states. The Schr\"{o}dinger bound state solution
is found as non-relativistic limit of the present model. It is noticed that
the solution with equal mixture of scalar and vector potentials can be
easily reduced into the well-known Schr\"{o}dinger solution for a particle
with an interaction potential field and a free field, respectively. We have
also studied the bound-state solutions for some special cases including the
non-relativistic limits (Schr\"{o}dinger equation for PHO and HO under
external magnetic and AB flux fields) and the KG equation for HO and PHO
interactions. The results show that the splitting is not constant and
dependent on the strength of the external magnetic field and AB flux field.

\acknowledgments We thank the referee(s) for the invaluable suggestions and
critics that have greatly helped in improving this paper. S.M. Ikhdair
acknowledges the support of the Scientific and Technological Research
Council of Turkey.

\appendix

\section{A Short Solution to KG Bound States}

Here, we seek to find a short solution to Eq. (9). By substituting Eqs. (3),
(5) and (6) into Eq. (9), we obtain a $2D$ Schr\"{o}dinger-type equation for
the anharmonic oscillator $V(r)=\gamma ^{2}r^{2}$ satisfying the radial wave
function $R(r),$ 
\begin{equation}
\left( \frac{d^{2}}{dr^{2}}-\frac{\left( \beta ^{2}-1/4\right) }{r^{2}}+\nu
^{2}-\gamma ^{2}r^{2}\right) R(r)=0,
\end{equation}%
where $g(r)=r^{-1/2}R(r).$ For brevity, the solution of the above equation
can be found through Eq. (17) of Ref. [5] by simply making the changes: 2$%
\mu E_{nL}^{\prime }/\hbar ^{2}\rightarrow \nu ^{2},$ 2$\mu B^{2}/\hbar
^{2}\rightarrow \gamma ^{2}$ and $2L+1\rightarrow 2\beta $ in Eq. (19) of
Ref. [5] and then finally obtain our Eqs. (12) and (13). A first inspection
on the asymptotic behavior of Eq. (A1), we find out that when $r$ approaches 
$0,$ the radial wave function $R(r)\sim r^{q},$\ $q=\beta +1/2>0$ and when $%
r\rightarrow \infty ,$ $R(r)\sim \exp \left( -\gamma r^{2}/2\right) $ and
both solutions are satisfying the finiteness of the radial wave function $%
g(r=0)=0$ and $g(r\rightarrow \infty )\rightarrow 0.$ Therefore, for the
entire range $r\in (0,\infty ),$ we consider the general solution $%
g(r)=r^{\beta }\exp \left( -\gamma r^{2}/2\right) h(r),$ $\beta >0,$ where $%
h(r)$ is the associated Laguerre polynomials, i.e., $%
h(r)=L_{n}^{(a)}(br^{2}).$ With these behaviors, the 3D wave function (20)
of Ref. [5] gives the 2D wave function as in Eq. (15) when one uses the
relation:%
\begin{equation*}
g(r)=Nr^{-1/2}R(r)=Nr^{L+1/2}\exp (-\sqrt{\frac{\mu }{2\hbar ^{2}}}%
Br^{2})L_{n}^{\left( L+1/2\right) }\left( \sqrt{\frac{2\mu }{\hbar ^{2}}}%
Br^{2}\right)
\end{equation*}%
\begin{equation}
=Nr^{\beta }\exp (-\frac{1}{2}\gamma r^{2})L_{n}^{\left( \beta \right)
}\left( \gamma r^{2}\right) ,
\end{equation}%
where $N$ is the normalization constant.

{\normalsize 
}

\bigskip

\baselineskip= 2\baselineskip
\bigskip \newpage

\bigskip

\begin{table}[tbp]
\caption{Specific values of the constants in the solution of Eq. (20).}%
\begin{tabular}{lll}
\tableline Constants: &  & ($\alpha _{3}=0$ case) \\ 
\tableline$\xi _{1}=\gamma ^{2}/4$ &  & $\xi _{2}=\nu ^{2}/4$ \\ 
$\xi _{3}=\beta ^{2}/4$ &  & $\alpha _{1}=1$ \\ 
$\alpha _{2}=\alpha _{3}=\alpha _{4}=\alpha _{5}=0$ &  & $\alpha _{6}=\xi
_{1}=\gamma ^{2}/4$ \\ 
$\alpha _{7}=-\xi _{2}=-\nu ^{2}/4$ &  & $\alpha _{8}=\xi _{3}=\beta ^{2}/4$
\\ 
$\alpha _{9}=\alpha _{6}=\gamma ^{2}/4$ &  & $\alpha _{10}=\beta +1$ \\ 
$\alpha _{11}=\gamma $ &  & $\alpha _{12}=\beta /2$ \\ 
$\alpha _{13}=-\gamma /2$ &  &  \\ 
\tableline &  & 
\end{tabular}%
\end{table}

\bigskip

\bigskip

\end{document}